\documentclass[12pt]{iopart}

\usepackage{graphicx}
\newcommand{\AaA}{{\it Astron. Astrophys.} }
\newcommand{\PRT}{{\it Phys. Rept.} }
\newcommand{\MNRAS}{{\it Mon. Not. R. Astron. Soc.} }
\newcommand{\ApJ}{{\it Astrophys. J.} }
\newcommand{\Nature}{{\it Nature} }
\newcommand{\ApJS}{{\it Astrophys. J. Supp.} }
\newcommand{\PRD}{{\it Phys. Rev. D.} }
\newcommand{\PLB}{{\it Phys. Lett. B.} }
\newcommand{\ICRC}{{\it International Cosmic Ray Conference} }
\newcommand{\NPB}{{\it Nucl. Phys. B.} }

\begin{document}
\title{The Galactic positron flux and dark matter substructures}

\author{Qiang Yuan and Xiao-Jun Bi} 

\address{Key Laboratory of Particle Astrophysics, Institute
of High Energy Physics, Chinese Academy of Sciences, P.O.Box 918-3,
Beijing 100049, P. R. China}

\ead{yuanq@ihep.ac.cn, bixj@ihep.ac.cn}

\begin{abstract}
{In this paper we calculate the Galactic positron flux from dark
matter annihilation in the frame of supersymmetry, taking 
the enhancement of the flux by existence of dark matter substructures
into account. The propagation of positrons in the Galactic magnetic
field is solved in a realistic numerical model GALPROP. 
The secondary positron flux is recalculated in the GLAPROP model.
The total positron flux from secondary products and dark matter annihilation
can fit the HEAT data well when taking a cuspy  density profile of the 
substrctures.
}
\end{abstract}
\pacs{95.35.+d} 
\noindent{\it Keywords}: dark matter, positron, propagation

\section{Introduction}
\label{intro}
The astronomical observations indicate that most 
of the matter in our universe is dark (see for example \cite{Jungman96}). 
The evidences come mainly from the gravitational effects of the dark
matter component, such as the rotation curves of
spiral galaxies \cite{Begeman91}, the gravitational
lensing \cite{Tyson95} and the dynamics of galaxy clusters \cite{White93}.
The studies of primordial nucleosynthesis 
\cite{Peebles71}, structure formation \cite{Davis85} and 
cosmic microwave background (CMB) \cite{Spergel03} show that this
so-called dark matter (DM) is mostly non-baryonic. 
The nature of the non-baryonic dark matter remains one of 
the most outstanding puzzles in particle physics and cosmology.

A large amount of theoretical models of DM
have been proposed in literatures \cite{Bertone04}.
All these candidates of non-baryonic DM particles
require physics beyond the standard model of particle physics.
Among the large amount of candidates, the most attractive
scenario involves the weakly interacting massive particles (WIMPs).
In particular, the minimal supersymmetric extension of the standard model
(MSSM) provides an excellent WIMP candidate as the lightest supersymmetric
particle (LSP), usually the neutralino, which are stable due to R-parity
conservation \cite{Jungman96}.

WIMPs are possible to be detected beyond the gravitational effects. 
One viable way is to detect the elastic scattering signals of DM particles
on the detector nuclei (direct searches) \cite{Goodman85,Bottino94,
Bernabei98,Bernabei99}. This is the most sensitive method
at present for DM detection \cite{Munoz04}.
Another way is to search the self-annihilation prodcuts 
of the DM particles (indirect searches), such as neutrinos
\cite{Barger02}, $\gamma$-rays \cite{Bergstrom01,deBoer04}, 
antiprotons \cite{Donato04} and positrons \cite{Baltz99,Kane02}. 
Since the positrons, antiprotons and diffuse
$\gamma$-rays are secondary particles in the cosmic rays with low
fluxes, these kinds of DM annihilation products are easier to be 
distinguished from the astrophysical background.
Direct and indirect detections of DM particles are complementary ways
to each other.

The balloon-borne instrument High-Energy Antimatter Telescope (HEAT)
was designed to determine the positron fraction over a wide energy 
range with high statitical and systematic accuracy. Based on 
the three flights(1994, 1995 and 2000), this collaboration reported
an excess of positrons with energies higher than several $GeV$ and
peaked in the range $7\sim 10\ GeV$ \cite{Barwick97,Coutu01}. It
seems that the astrophysical production cannot give enough positrons
as observed \cite{Moskalenko98}, and there might exist exotic
sources of positrons. It has been pointed out that the excess may indicate
the signal of DM annihilation \cite{Baltz99, Kane02}. 
However, Baltz \etal showed that for supersymmetry (SUSY) DM 
a ``boost factor'' $\ge 30$ has to be introduced to fit the HEAT data 
\cite{Baltz02}. It was thought that the local clumpiness of DM
distribution might account 
for this ``boost factor''. However, Hooper \etal argued that it was less
possible for a DM clump to be close enough to the Earth to contribute
sufficient positron flux \cite{Hooper04}. 

Stimulated by the recent numerical simulation which shows that
the lightest subhalos in the Galaxy could be extended to $10^{-6} M_{\odot}$
with a huge amount of about $10^{15}$ \cite{Diemand05}, we try
to calculate the Galactic positron flux again in the frame of supersymmetry
taking the enhancement by the large amount of subhalos into account.
There was effort try to explain the HEAT data by the nearby mini-clumps
\cite{Cumberbatch06}. However, the possibility of this scenario is also
extremely low \cite{Lavalle06}. Another difference of this work from 
the previous studies most of which adopt the 
Green's function method to calculate the positron's propagation
is that we calculate its propagation in the Galactic magnetic field 
using the numerical package GALPROP \cite{Strong98}, which adopts
realistic distribution of interstellar gas and radiation field.
Our result shows that the two methods lead to some difference.
To give a consistent result we also calculate the secondary positron
flux using GALPROP. The propagation parameters are adjusted to fit
observations such as B/C, proton and electron spectra. Our results
show that the secondary positron has better agreement with the HEAT
data than which adopted in previous studies \cite{Baltz99,Kane02,
Baltz02,Cumberbatch06} and decreases the tension between predictions
and data.

The paper is orgnized as following: in Sec. \ref{anni} we present the SUSY
model. In Sec. \ref{profile} we give our treatment of subhalos distribution and
density profiles. Sec. \ref{galprop} introduces the positron's propagation and
the GALPROP model. The results are given in Sec. \ref{result}.

\section{Positron Production from Neutralino Annihilation}
\label{anni}

The LSP, generally the
lightest neutralino, is stable in the R-parity conservative MSSM and
provides a natural candidate of dark matter.  It is
a combination state of the supersymmetric partners of the photon, $Z^0$
boson and neutral Higgs bosons. 

Positrons can be produced in several neutralino annihilation modes. Most
of them come from the decay of gauge bosons produced in chanels $\chi\chi
\rightarrow ZZ,\ \chi\chi\rightarrow W^+W^-$, or from the cascades 
of final particles, such as fermions and Higgs bosons. The 
spectrum of positrons depends on the neutralino mass and its
annihilation modes. There is 
also a direct channel $\chi\chi\rightarrow e^+e^-$ and 
produces monochrome positrons with energy $E_{e^+}= m_{\chi}$.
However, the branching 
ratio of this channel is usually small. In the following discussion
we neglect this ``line'' contribution to the positron spectrum. 

The source function of positrons from DM annihilation can be written as
\begin{equation}
Q(E_{e^+},{\bf r})=\frac{\langle\sigma v\rangle}{2m_{\chi}^2}\frac{{\rm d}n}{{\rm d}E}
\rho^2({\bf r}),
\label{positronsource}
\end{equation}
where $\sigma$ is the positron generating cross-section, 
${{\rm d}n}/{{\rm d}E}$ is the positrons spectrum in one annihilation 
by a pair of neutralinos and
$\rho({\bf r})$ is neutralino density distribution in space.
The source term $Q(E_{e^+},{\bf r})$ is given in
unit of $GeV^{-1}m^{-3}s^{-1}$.

The source term is calculated in MSSM by doing a random scan 
using the software package DarkSUSY \cite{Gondolo00}. In these SUSY
models we choose a few models which satisfy all the experimental bounds
and give large fluxes and appropriate spectrum so that we can fit 
the HEAT data well.
However, there are more than one hundred free SUSY breaking
parameters even for the R-parity conservative MSSM.
A general practice is to assume some relations between the parameters
and greatly reduce the number of free parameters.
For the processes related with dark matter production and annihilation,
only seven parameters are relevant under some simplifying assumptions,
i.e., the higgsino mass parameter $\mu$, the wino mass parameter $M_2$,
the mass of the CP-odd Higgs boson $m_A$, the ratio of the Higgs
Vacuum expectation values $\tan\beta$, the scalar fermion mass parameter
$m_{\tilde{f}}$, the trilinear soft breaking parameter $A_t$
and $A_b$. To determine the low energy spectrum of the SUSY particles
and coupling vertices,
the following assumptions have been made: all the sfermions
have common soft-breaking mass parameters $m_{\tilde{f}}$; all trilinear
parameters are zero except those of the third family; the bino, wino, and 
gluino have the
mass relations, $M_1=5/3\tan^2\theta_W M_2$, $M_3=\alpha_3(M_Z)/
{\alpha_{\mbox{em}}}\sin^2\theta_W M_2$,
coming from the unification
of the gaugino mass at the grand unification scale.
                                                                               
The parameters are constrained in the following ranges:
$50 GeV < |\mu|,\ M_2,\ M_A,\ m_{\tilde{f}} < 10 TeV$;
$1.1 < \tan\beta < 55$; $-3m_{\tilde{q}} < A_t, \ A_b < 3m_{\tilde{q}}$;
$\mbox{sign}(\mu)=\pm 1$.
The SUSY models are required to satisfy the theoretical consistency
requirement, such as the correct vacuum breaking pattern,
the neutralino being the LSP and so on. The accelerator data
constrains the models further
from the spectrum requirement, the invisible Z-boson width,
the branching ratio of $b\to s\gamma$ and the muon magnetic moment.

The constraint from cosmology is also taken into account
by requiring the relic abundance of neutralino
$0.091 < \Omega_\chi h^2< 0.118 $,
which corresponds to the 3$\sigma$ bound from
the cosmological observations \cite{Spergel03}. 
The effect of coannihilation between the fermions is taken into account
when calculating the relic density numerically.

\section{Dark Matter Distribution}
\label{profile}

To determine the positron source term from DM annihilation in
Eq. (\ref{positronsource}) we have to specify the DM density profile 
$\rho({\bf r})$. 
Based on the N-body simulation results the DM density profile
can be written in a general form as \cite{Zhao96}
\begin{equation}
\rho=\frac{\rho_s}{(r/r_s)^{\gamma}[1+(r/r_s)^{\alpha}]^{(\beta
-\gamma)/\alpha}},
\label{rho}
\end{equation}
where $\rho_s$ and $r_s$ are the scale density and scale radius 
respectively. These two parameters are determined by the measurement
of the virial mass of the halo and the concentration parameter determined
by simulation. 
The NFW profile was first proposed by Navarro, Frenk and White
\cite{Navarro97} with $(\alpha,\beta,\gamma)=(1,3,1)$.
However, Moore \textit{et al.} gave another form of DM profile
with $(\alpha,\beta,\gamma)=(1.5,3,1.5)$ to fit their simulation
result \cite{Moore98}. The Moore profile has steeper slope
near the Galactic center than the NFW profile. 
There are also simulations showing that
the density profile may not be universial \cite{Jing00,Reed05}.
Reed et al. show that $\gamma = 1.4 - 0.08\log(M/M_*)$ increases for halos
with smaller masses \cite{Reed05}. Therefore we also adopt a $\gamma = 1.7$ 
profile for the whole range of DM halos for simplicity.
At present there are controversials on which one represents the 
actual density profile. 

\begin{figure}[!ht]
\resizebox{13.cm}{!}{\includegraphics[scale=1]{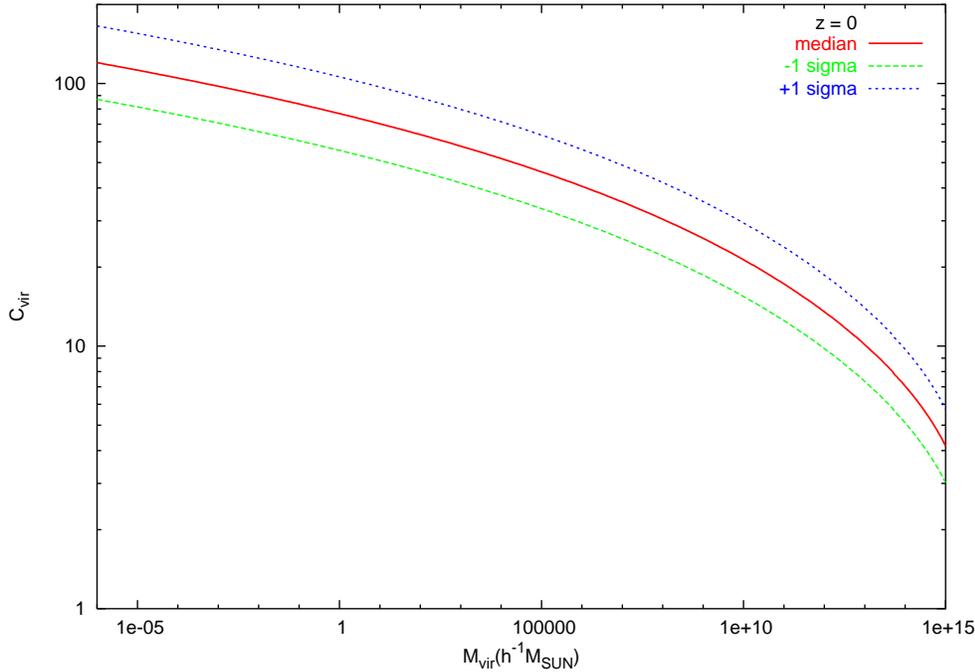}}
\caption{$c_{\mbox{vir}}$ as a function of subhalo mass $m_{\mbox{sub}}$
predicted by the Bullock model \cite{Bullock01}. The median and 
the 1$\sigma$ values are given.}
\label{con}
\end{figure} 

\begin{figure}[!ht]
\resizebox{13.cm}{!}{\includegraphics[scale=1]{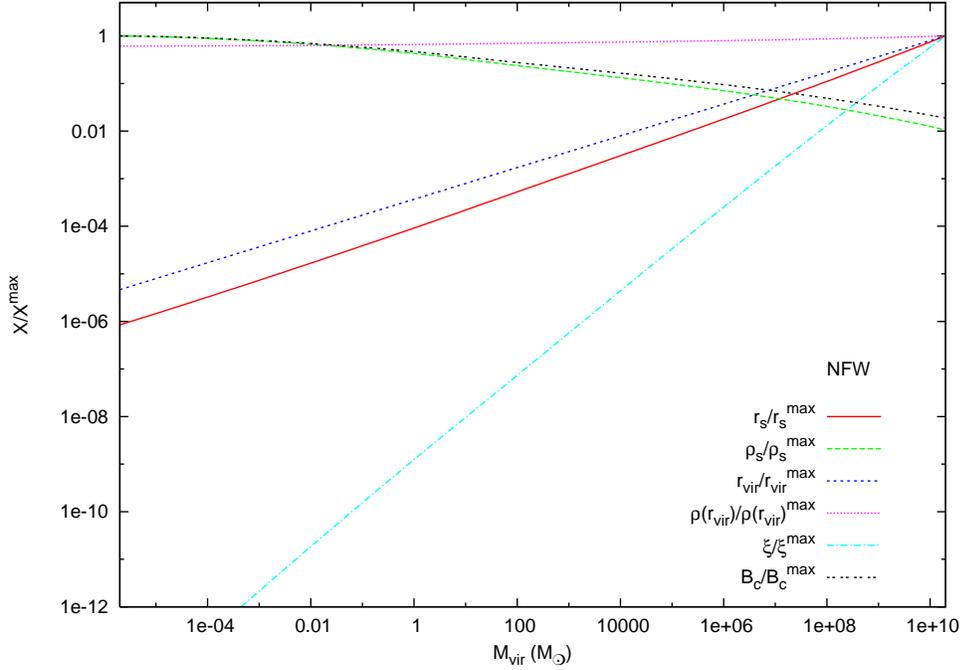}}
\caption{Some characteristic quantities of DM halos with NFW mass 
distribution as functions of masses.}
\label{clump_para}
\end{figure}

For a DM halo with mass $M_{\mbox{vir}}$, we adopt a semi-analytic model
of Bullock \etal \cite{Bullock01}, which describes the 
concentration parameter $c_{\mbox{vir}}$ as a function of viral mass and 
redshift, to determine the parameters in the density profile. 
We plot the $c_{\mbox{vir}}-M_{\mbox{vir}}$ relation in Figure
\ref{con}, with the $1\sigma$ range. The median $c_{\mbox{vir}}
-M_{\mbox{vir}}$ relation at redshift zero is adopted. The scale radius 
is then determined as 
$r^{\mbox{nfw}}_{\mbox{s}}=r_{\mbox{vir}}/c_{\mbox{vir}}$,
$r^{\mbox{moore}}_{\mbox{s}}=r^{\mbox{nfw}}_{\mbox{s}}/0.63$ and 
$r^{\gamma}_{\mbox{s}}=r^{\mbox{nfw}}_{\mbox{s}}/(2-\gamma)$. The virial
radius $r_{\mbox{vir}}$ in above formula is defined by 
$M_{\mbox{vir}}=4\pi\Delta\rho_{\mbox{c}}r_{\mbox{vir}}^3/3$ 
with $\Delta=200$ and the critical density of the universe 
$\rho_{\mbox{c}}=139M_{\odot}/kpc^3$. Once $r_{\mbox{s}}$ is fixed,
the scale density $\rho_{\mbox{s}}$ can be got by solving the equation
$M_{\mbox{vir}}=\int \rho({\bf r}){\rm d}^3{\bf r}$.
In Figure \ref{clump_para} we show the scale radius $r_{\mbox{s}}$, 
scale density $\rho_{\mbox{s}}$,
virial radius $r_{\mbox{vir}}$, surface mass density $\rho
(r_{\mbox{vir}})$, effective volume $\xi=\int (\rho/\rho_0)^2
{\rm d}^3{\bf r}$, intrinsic boost factor $B_{\mbox{c}}=
\frac{\int (\rho/\rho_0)^2{\rm d}^3{\bf r}}
{\int (\rho/\rho_0){\rm d}^3{\bf r}}$ as functions of masses
$M_{\mbox{vir}}$, for the DM halos with NFW mass distributions. 
We adopt a reference mass density $\rho_0=0.3GeV/cm^3$.
Each of these charicteristic quantities is scaled by 
its maximum value in the mass range $2\times10^{-6}\sim 2\times 
10^{10}M_{\odot}$, the minimal and maximal masses of the Milky Way
(MW) subhalos we adopt in the following discussion. 
We list these scale factors as: $r_{\mbox{s}}
^{\mbox{max}}=2.72kpc,\ \rho_{\mbox{s}}^{\mbox{max}}=3.52\times
10^9M_{\odot}/kpc^3,\ r_{\mbox{vir}}^{\mbox{max}}=55.62kpc,
\ \rho(r_{\mbox{vir}})^{\mbox{max}}=1.52\times10^{-4}GeV/cm^3,
\ \xi^{\mbox{max}}=1909.11kpc^3,\ B_{\mbox{c}}^{\mbox{max}}=39.89.$
The behaviors of these quantities can be understood as following.
For NFW profile, an analytic calculation shows that 
$r_{\mbox{vir}}\propto M_{\mbox{vir}}^{1/3}$, and
$r_{\mbox{s}}\propto M_{\mbox{vir}}^{1/3}/c_{\mbox{vir}}$. Because the
concentration parameter $c_{\mbox{vir}}$ is weakly dependent on the
mass according to Figure \ref{con}, $r_{\mbox{s}}$ is approximately 
proportional to $M_{\mbox{vir}}^{1/3}$. The scale density $\rho_{\mbox{s}}$ 
is nearly $\propto c_{\mbox{vir}}^3/\ln(1+c_{\mbox{vir}})$, which shows a
slight decrease as the increase of halo mass, while the surface density
$\rho(r_{\mbox{vir}})\sim\rho_{\mbox(s)}/c_{\mbox{vir}}^3\propto 1/
\ln(1+c_{\mbox{vir}})$ is almost constant. The effective volume $\xi$ 
is approximately scaled with $M_{\mbox{vir}}c_{\mbox{vir}}^3$, and the 
intrinsic boost factor $B_{\mbox{c}}\propto c_{\mbox{vir}}^3$. 
$B_{\mbox{c}}$ is a little smaller for higher mass halos, because it is 
less concentrated for higher mass halo.

Having known the general properties of the DM halos given above,
we now turn to the discussion of the MW subhalos.
High resolution simulations have revealed that a large number of self-bound
substructures survived in the galactic halos \cite{Tormen98,Klypin99,
Moore99,Ghigna00,Springel01,Zentner03,Delucia04,Kravtsov04}. 
Especially a recent simulation conducted by Diemand
et al. \cite{Diemand05} showed that the first generation objects as
light as the Earth mass can survive until today. The number of such
minihalos is huge, reaching $10^{15}$.
The existence of a wealth of
subhalos will greatly enhance the DM annihilation flux, since the
annihilation rate is propotional to the density square as shown
in Eq. (\ref{positronsource}). 

The simulations show that the number
density of subhalos can be approximately given by an isothermal
profile with a core \cite{Diemand04}
\begin{equation}
\label{dis}
n(r)=2n_H(1+(r/r_H)^2)^{-1}\ ,
\end{equation}
where $n_H$ is the relative number density at the scale
radius $r_H$, with $r_H$ being about $0.14$ times the halo virial radius
$r_H=0.14r_{\mbox{vir}}$. 
The result given above agrees well with that in another recent
simulation by Gao et al.\cite{Gao04}.
Eq. (\ref{dis}) shows that the radial distribution of 
substructures is generally shallower than the density profile 
of the smooth background in Eq. (\ref{rho}).
This is due to the tidal disruption of substructures
which is most effective near the galactic center.

The mass function is well fitted to the simulation result as 
${\rm d}n/{\rm d}m_{\mbox{sub}}\propto m_{\mbox{sub}}^{-1.9}$.
We then get the number density of 
substructures with mass $m_{\mbox{sub}}$ at the position 
$r$ to the galactic center
\begin{equation}
\label{prob}
n(m_{\mbox{sub}},r)=n_0 \left(\frac{m_{\mbox{sub}}}
{M_{\mbox{vir}}}\right)^{-1.9} (1+(r/r_H)^2)^{-1}\ ,
\end{equation}
where $M_{\mbox{vir}}$ is the virial mass of the MW, 
$n_0$ is the normalization factor determined by requiring the
number of subhalos with mass larger than $10^8M_{\odot}$ is about 500
in a halo with $M=2\times 10^{12}M_{\odot}$\cite{Moore99}.

Taking the subhalos into account the density square in Eq. 
(\ref{positronsource}) is given by
\begin{equation}
\rho^2({\bf r}) \to \langle \rho^2({\bf r}) \rangle
= \rho^2_{\mbox {smooth}}({\bf r}) + \langle \rho^2_{\mbox{sub}}
({\bf r}) \rangle, 
\end{equation}
where $\langle \rho^2_{\mbox{sub}}({\bf r}) \rangle$ means the average
density square of subhalos according to the distribution probability.
Since there is no correlation between the mass and spatial distribution
in Eq. (\ref{prob}) we get $\langle \rho^2_{\mbox{sub}}({\bf r}) \rangle$
at the position ${\bf r}$ is given by an integral of mass
\begin{eqnarray}
\langle \rho^2_{\mbox{sub}}({\bf r}) \rangle &=&
\int_{m_{\mbox{min}}}^{m_{\mbox{max}}} n(m_{\mbox{sub}},r)
\left( \int \rho_{\mbox{sub}}^2 {\rm d}V \right)
\cdot {\rm d}m_{\mbox{sub}} \nonumber\\ 
&=& \frac{n_0}{1+(r/r_H)^2} \int_{m_{\rm min}}^{m_{\rm max}} 
f(m_{\mbox{sub}}) \cdot {\rm d}m_{\mbox{sub}}\ \ ,
\label{averrhosq}
\end{eqnarray}
where $\rho_{\mbox{sub}}$ refers to density of the subhalo at ${\bf r}$ and 
$V$ is its volume.
We defined a function $f(m_{\mbox{sub}})$ in the equation above as
\begin{equation}
f(m_{\mbox{sub}})=\left (\frac{m_{\mbox{sub}}}{M_{vir}}\right )^{-1.9}
\int_0^{r_{\mbox{\tiny vir}}(m_{\mbox{\tiny sub}})}
4\pi {r^{\prime}}^2\rho^2_{\mbox{sub}}(r^{\prime}) {\rm d}r^{\prime},
\label{fm}
\end{equation}
which depends only on the subhalo mass since the virial
radius and the density profile are all determined by the 
subhalo mass.
The minimal subhalos can be as light as 
$10^{-6} M_{\odot}$ \cite{Diemand05}
while the maximal mass of substructures is taken 
to be $0.01 M_{\mbox{vir}}$ \cite{Bi06}.

We notice that there are unphysical
singularities at the center of the subhalo which may lead to
$f(m_{\mbox{sub}})$ divergent for the index $\gamma$ equal to or
greater than $1.5$.
A cutoff $r_{\mbox{cut}}$ is introduced within which the DM 
density is kept a constant value $\rho_{\mbox{max}}$ due to the balance
between the annihilation rate and the rate to fill the region
by infalling DM particles \cite{Berezinsky92}. Applying typical setting
of parameters, we get $\rho_{\mbox{max}}\sim 10^{18} - 10^{19} 
M_\odot/kpc^3$. 
In the following discussion, $\rho_{\mbox{max}}=10^{18}M_{\odot}/
kpc^3$ is adopted for NFW and Moore profiles while we consider
$\rho_{\mbox{max}}=10^{18} - 2\times 10^{19}M_\odot/kpc^3$ for the
profile with $\gamma=1.7$. 

\begin{figure}[!ht]
\resizebox{13.cm}{!}{\includegraphics[scale=1]{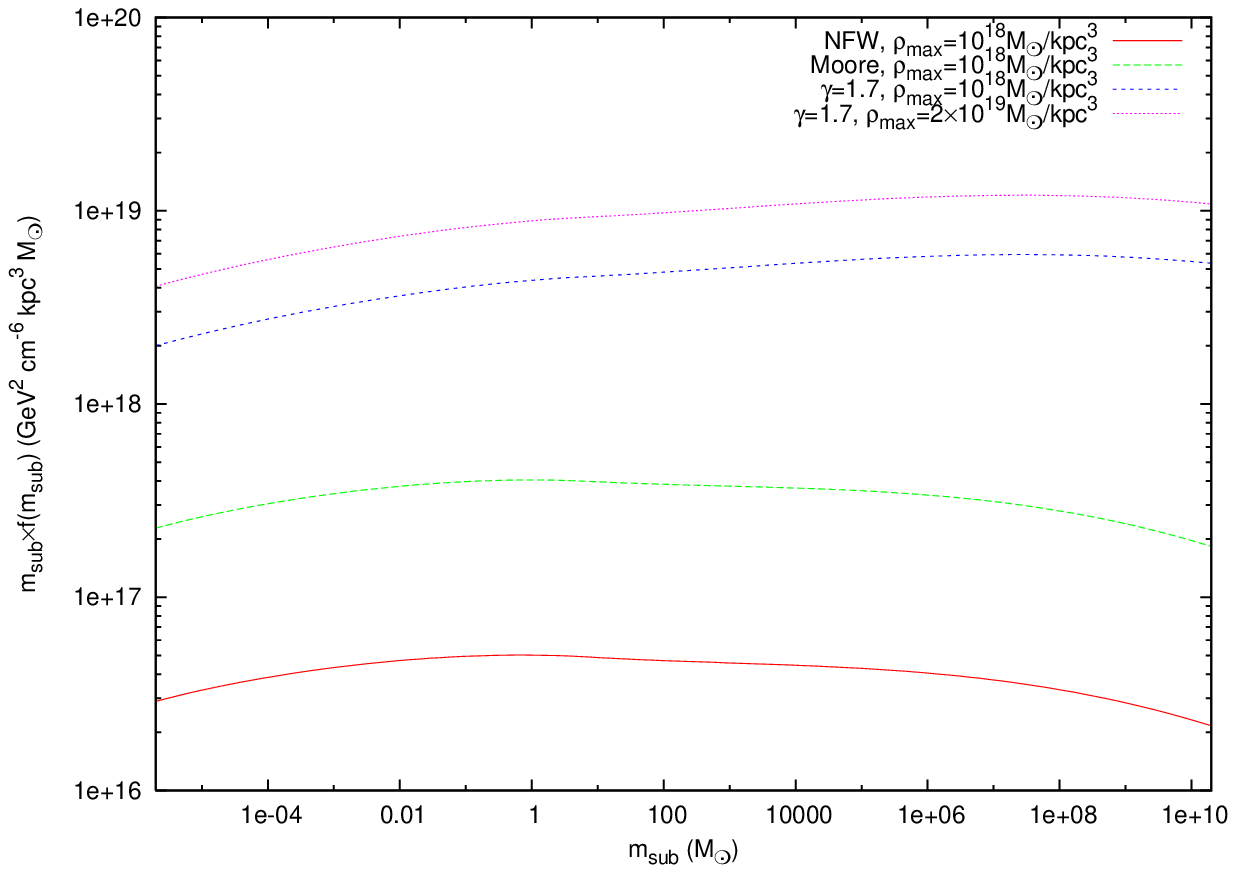}}
\caption{$m_{\mbox{sub}}\times f(m_{\mbox{sub}})$ as a function of subhalo 
mass $m_{\mbox{sub}}$.}
\label{mfm}
\end{figure}

From Eq. (\ref{averrhosq}) we know that, $\langle\rho^2_{\mbox{sub}}
\rangle\propto\int f(m_{\mbox{sub}})\,{\rm d}m_{\mbox{sub}}
=\int m_{\mbox{sub}}f(m_{\mbox{sub}})\,{\rm d}\ln m_{\mbox{sub}}$.
The quantity $m_{\mbox{sub}}\times f(m_{\mbox{sub}})$ is plotted in 
Figure \ref{mfm}, which shows the relative contribution to the 
annihilation signals from subhalos at different mass bins at 
the logarithmic scale.
We can see from this figure that subhalos of different 
mass bins contribute almost the same to the annihilation products.
By extending the minimal 
subhalo mass from the present numerical resolution of about $10^6 M_\odot$ 
to $10^{-6} M_\odot$, the contribution will be $\sim 4$ times larger.

\begin{figure}[htb]
\resizebox{13.cm}{!}{\includegraphics{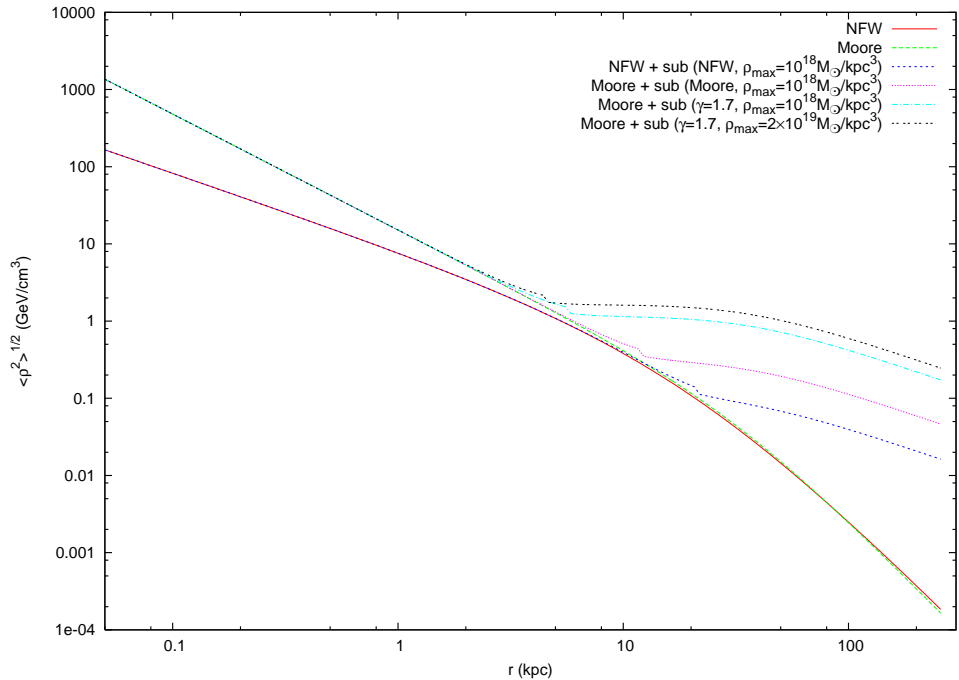}}
\caption{$\langle{\rho^2(r)}\rangle^{1/2}$ distribution of Galaxy 
for various DM profiles.}
\label{rho2}
\end{figure} 
 
Figure \ref{rho2} shows $\langle{\rho^2}\rangle^{1/2}$ for different DM 
profiles. The existence of subhalos will increase $\langle{\rho^2}
\rangle^{1/2}$ significantly at large radii. Because the annihilation
source is density square dependent, the further clumpiness of DM will
effectively increase the annihilation signals, and can play the role of 
a ``boost factor''. The more concentrated of the halo profile, the larger
of the $\langle{\rho^2}\rangle^{1/2}$ value. 
While at small radii there is no enhancement compared
to the smooth component for the tidal disruption makes the subhalos be
difficult to survive in inner Galaxy. We can see that the model with the 
most cuspy subhalos adopted here has $\sim 3$ times larger 
$\langle{\rho^2}\rangle^{1/2}$ than the smooth one at the solar 
location, which means that the positron flux observed on the Earth will 
be higher by about an order of magnitude.

\section{The Positron Propagation}
\label{galprop}
A complexity in calculating the positron flux is that positrons
are scattered by the Galactic magnetic field (GMF) and we have to calculate
its propagation in GMF.
The propagation of charged particles in the GMF
is diffusive. In addition they will also experience energy loss
processes due to ionization and Coulomb interaction in the interstellar
medium; for electrons and positrons
there are additional synchrotron radiation in the GMF,
bremsstrahlung radiation in the interstellar medium
and inverse Compton scattering in the interstellar
radiation field. The interaction between cosmic-ray particles and
the interstellar medium (mostly Hydrogen and Helium) will lead to the
fragmentation of nuclei. For radioactive nuclei the decay should be
also taken into account. 

The full propagation equation including the convection of
cosmic-ray particles and reacceleration processes 
is written in the form
\begin{eqnarray}
\frac{\partial \psi}{\partial t} &=&Q({\bf x},p)+\nabla\cdot(D_{xx}\nabla
\psi-{\bf V}\psi)+\frac{\partial}{\partial p}p^2D_{pp}\frac{\partial}
{\partial p}\frac{1}{p^2}\psi \nonumber \\
 &-& \frac{\partial}{\partial p}\left[\dot{p}\psi
-\frac{p}{3}(\nabla\cdot{\rm V}\psi)\right]-\frac{\psi}{\tau_f}-
\frac{\psi}{\tau_r},
\label{prop}
\end{eqnarray}
where $\psi$ is the density of cosmic ray particles per unit momentum interval,
$Q({\bf x},p)$ is the source term, $D_{xx}$ is the spatial diffusion
coefficient, ${\bf V}$ is the convection velocity. The third term
describes the reacceleration
using diffusion in momentum space. $\dot{p}$ is the momentum loss
rate, $\tau_f$ and $\tau_r$ are timescales for fragmentation and
radioactive decay respectively. 
                                           
The interaction between cosmic-ray particles and
the interstellar medium produces the secondary particles, such as
positrons, antiprotons and some other heavy elements.
The source term of the secondary cosmic-rays is given by
\begin{equation}
Q({\bf x},p)=\beta c\psi_p({\bf x},p)[\sigma_{H}(p)n_{H}({\bf x})+
\sigma_{He}(p)n_{He}({\bf x})],
\label{secsource}
\end{equation}
where $\psi_p({\bf x},p)$ is the density of primary cosmic-rays,
$\beta c$ is the velocity of particles, $\sigma_H$ and $\sigma_{He}$
are the production
cross sections for the secondary particles from the progenitor on H
and He targets, $n_H$ and $n_{He}$ are the interstellar Hydrogen
and Helium number densities, respectively.

The propagation equation (\ref{prop}) can be solved 
analytically under some simplification assumptions \cite{Baltz99,
Maurin01}. A numerical solution to Eq. (\ref{prop}) is given in the
GALPROP model developed by Strong and Moskalenko \cite{Strong98},
which takes all the relevant processes into account.
The realistic distributions
for the interstellar gas and radiation fields are adopted in GALPROP.
The detail of this model can be found in \cite{Strong98}.

The secondary-primary ratio such as B/C depends on the propagation 
process. Therefore the propagation parameters are adjusted to 
describe the B/C ratio, the electron and proton spectra \etal. 
We include the nuclei up to $Z=28$ and relevant isotopes.
The injection spectra of protons and heavier nuclei are assumed
to have the same power-law form in rigidity.
The nuclei injection index below and above the reference rigidity
at 9 GV are taken as $1.98$ and $2.42$ respectively. The electron
injection index below and above
4 GV are taken as $1.60$ and $2.54$ respectively.
For propagation, we use the diffusion
reacceleration model \cite{Moskalenko02}. 
The spatial diffusion coefficient is taken as $\beta D_0(\rho/\rho_0)^
\delta$, where $D_0=5.4\times 10^{28}$ cm s$^{-1}$, $\rho_0=4$ GV, and
$\delta=0.33$. The Alfv\'en speed is $v_A=30$ km
s$^{-1}$. The height of the propagation halo is taken as $z_h=4$ kpc.

\begin{figure}[!ht]
\resizebox{13.cm}{!}{\includegraphics{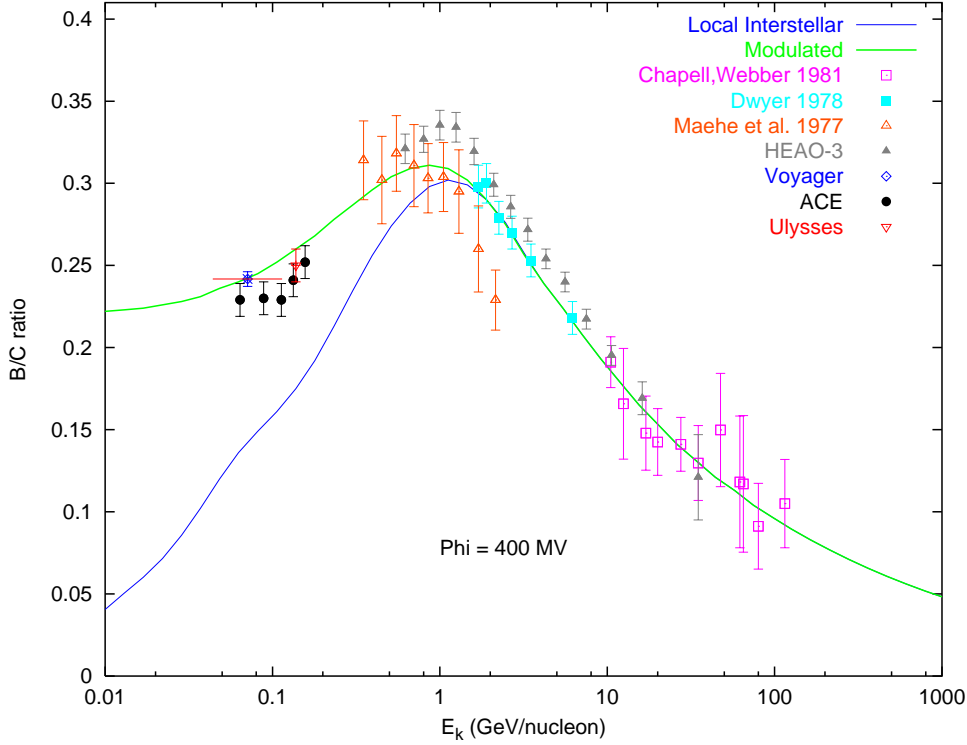}}
\caption{Calculated Boron-to-Carbon ratio compared with observational
data. The lower line represents the local interstellar value, while the 
upper one is the result after solar modulation with $\Phi=400MV$.
Observational data: ACE \cite{Davis00}, Ulysses \cite{DuVernois96},
Voyager \cite{Lukasiak99}, HEAO-3 \cite{Engelmann90}, for others
please refer to \cite{Stephens98}.}
\label{BC}
\end{figure}

In Figure. \ref{BC} we show the expected B/C from the propagation
model. The propagation model describes the observational data very well.
We will use this model to calculate the secondary positron spectrum
from cosmic rays fragmentation given in Eq. (\ref{secsource})and the 
primary positron propagation from DM annihilation given in 
Eq. (\ref{positronsource}).

\section{Results and Discussion}
\label{result}
\begin{figure}[ht]
\resizebox{13.cm}{!}{\includegraphics{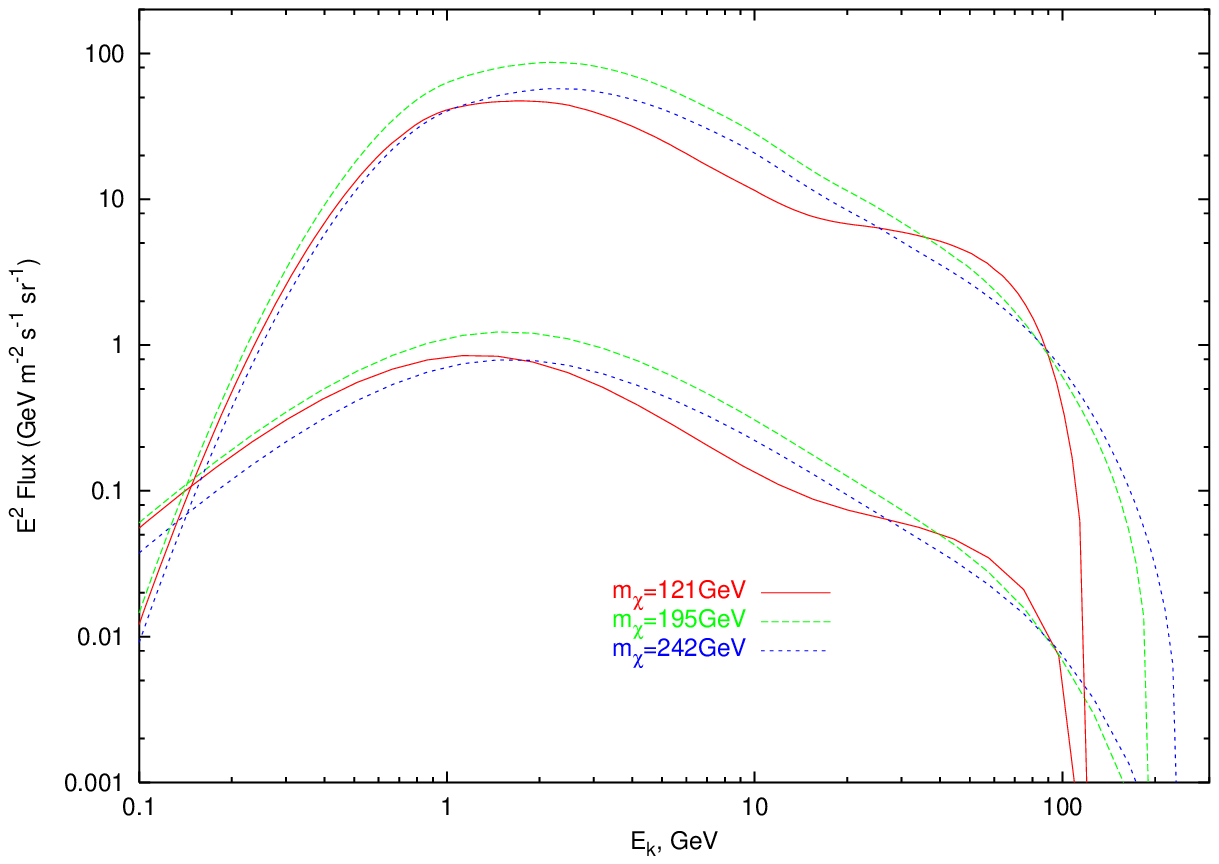}}
\caption{The propagated positron fluxes on the
Earth for three neutralino models. Moore profile, containing 
subhalos with $\gamma=1.7$ and $\rho_{\rm max}=2\times 10^{19} 
M_\odot/kpc^3$ is adopted. Upper lines show the local annihilation 
sources $E^2\,Q(E,{\bf r})$ in unit of GeV $m^{-3} s^{-1}$
for the three neutralino models.}
\label{spectrum}
\end{figure}
\begin{figure}[ht]
\resizebox{13.5cm}{!}{\includegraphics{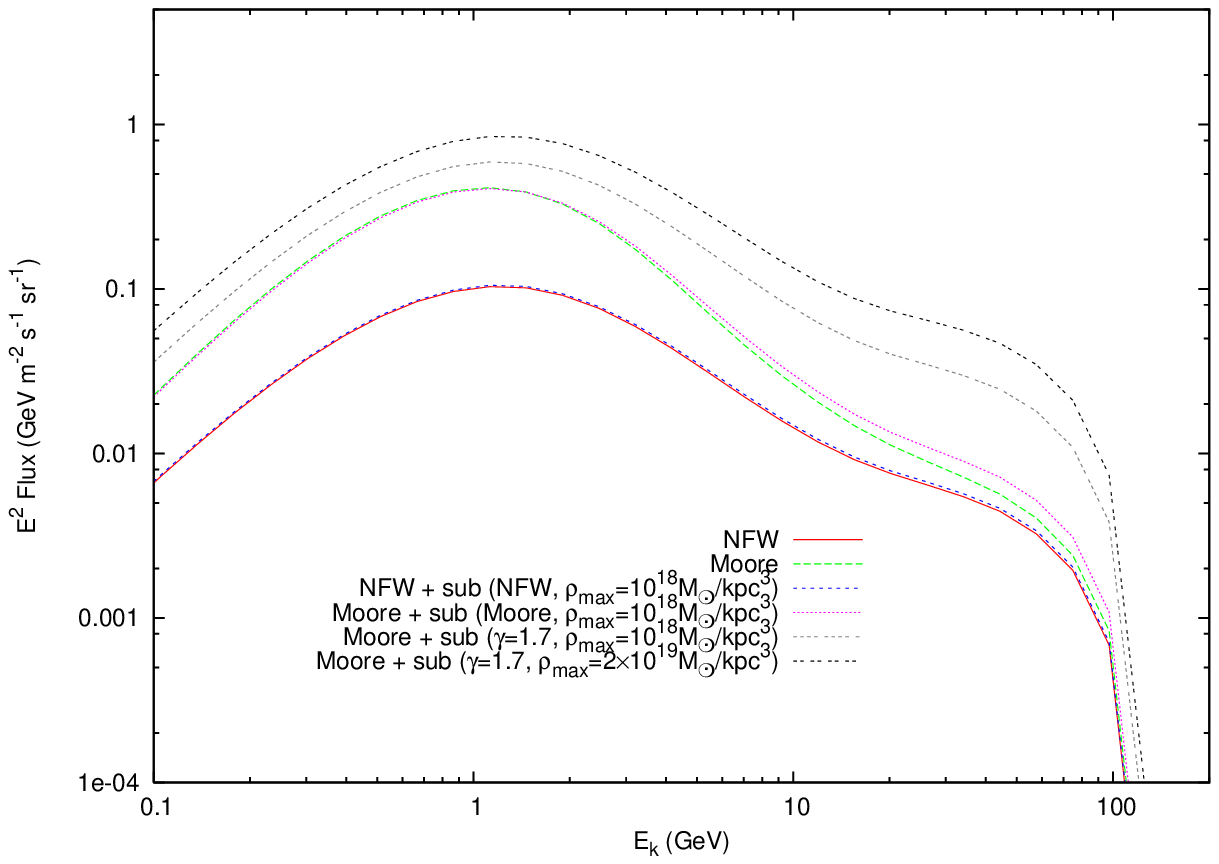}}
\caption{Positron flux on Earth for various DM profiles. The mass 
of neutralino is $121GeV$.}
\label{spectrum_prof}
\end{figure}

Once the SUSY model is specified and the contribution from subhalos
is taken into account we can incorporate the source term of primary
positrons in GALPROP and calculate its spectrum on the Earth.
In Figure \ref{spectrum} we show the propagated
positron fluxes for three different SUSY models with $m_\chi = 121, 
195, 242$ GeV. For comparison the local source terms $E^2Q(E,{\bf r}= {\bf r}_\odot)$ 
according to Eq. (\ref{positronsource}) 
are also shown. We see that propagation makes the spectra
some different from the source spectra. This is because the propagation
is energy dependent.
The source spectra are somewhat different among the three models
because these models lead to different annihilation final states.
In the calculation a Moore profile for the smooth 
component and $\gamma=1.7$ profile for subhalos with $\rho_{\rm max}
=2\times 10^{19} M_\odot/kpc^3$ are adopted. 

In Figure \ref{spectrum_prof} we show the positron flux for the
$m_{\chi}= 121GeV$ model for different DM profiles.
It indicates that the fluxes can be different for
several times among various DM profiles.
The cuspier profiles lead to higher positron fluxes.
It should be noted that because of the propagation effects the
spectra from different DM profiles are different, although they
have the same source spectrum.
At lower energy the fluxes have bigger differences, since
the low energy positrons have longer propagation distance, and may
trace the $\langle\rho^2\rangle$ farther away from the Earth.
The differences between the NFW (or Moore) profile with and without 
subhalos are not significant. For subhalos with $\gamma =1.7$ profile 
the flux is greatly enhanced.

\begin{figure}
\resizebox{13.cm}{!}{\includegraphics{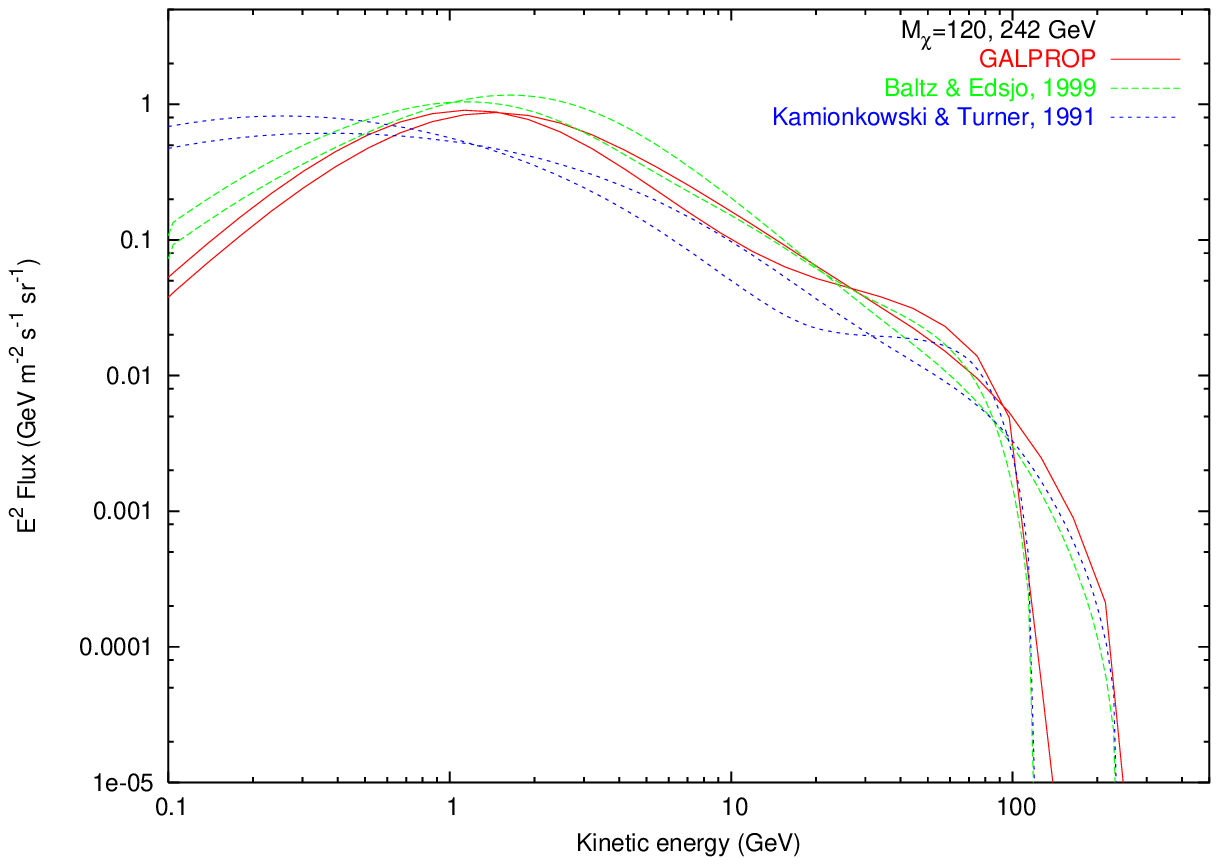}}
\caption{Comparison of positron fluxes of different propagation models.}
\label{propmodel}
\end{figure}

We then compared the results between GALPROP and other propagation methods.
The results are shown in Figure \ref{propmodel}.
The fluxes for $m_\chi =121$ and $242$ GeV
based on the Green's functions given by
Kamionkowski \& Turner \cite{Kamionkowski91} and Baltz \& Edsjo 
\cite{Baltz99} are plotted. The BE Green's function is to solve the
diffusive propagation equation analytically and similar to 
the GALPROP method. However, 
GALPROP uses realistic astrophysical inputs, and can contain more
complex physical process such as reacceleration, while in BE's method
only an average energy loss rate is considered and without convection
or reacceleration. After adjusting the propagation parameters to be 
equivalent, we get similar results between these two models. 
KT's result shows a larger difference from these two models. This is
comprehensible that the KT 
Green's function is acquired by solving the leaky-box equation of
propagation with a position-independent source.
At low energies the discrepancy is larger
because of longer propagation path.

\begin{figure}[htb]
\resizebox{15.5cm}{!}{\includegraphics{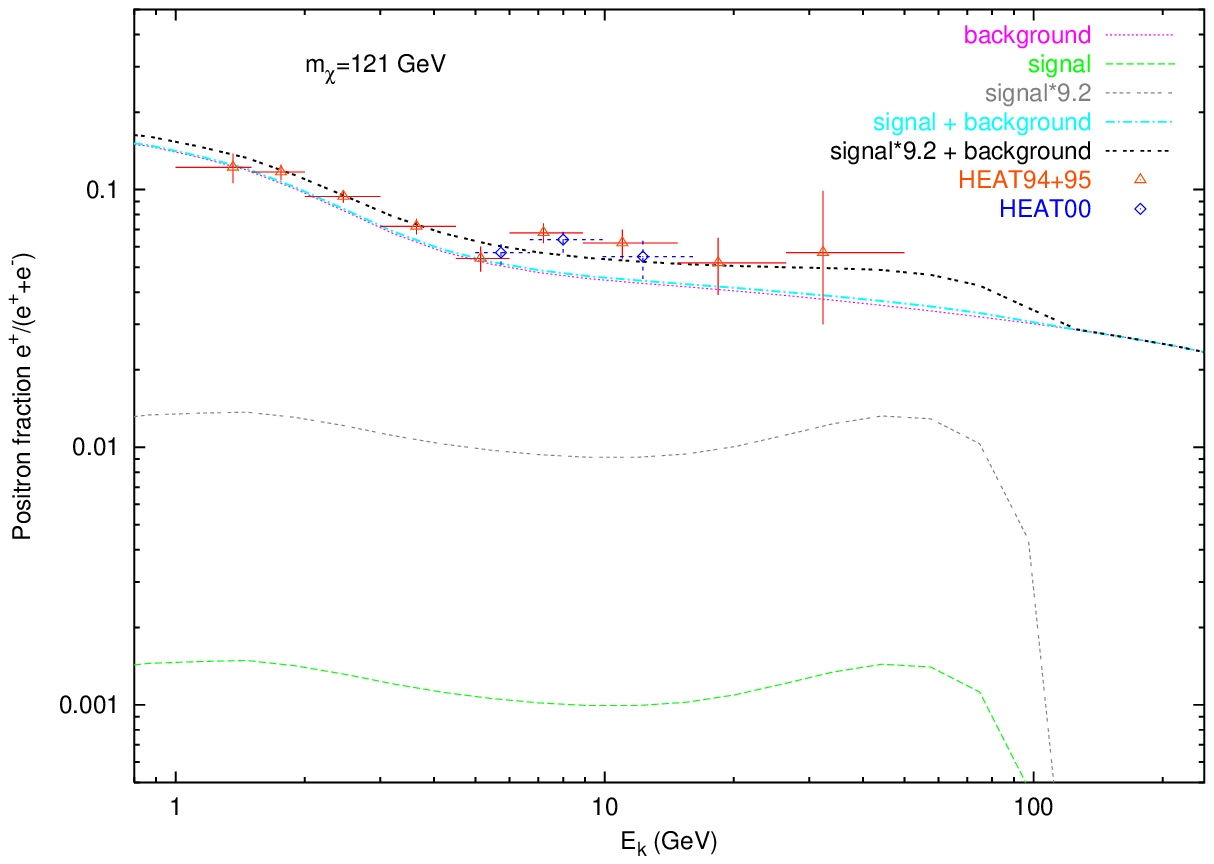}
\includegraphics{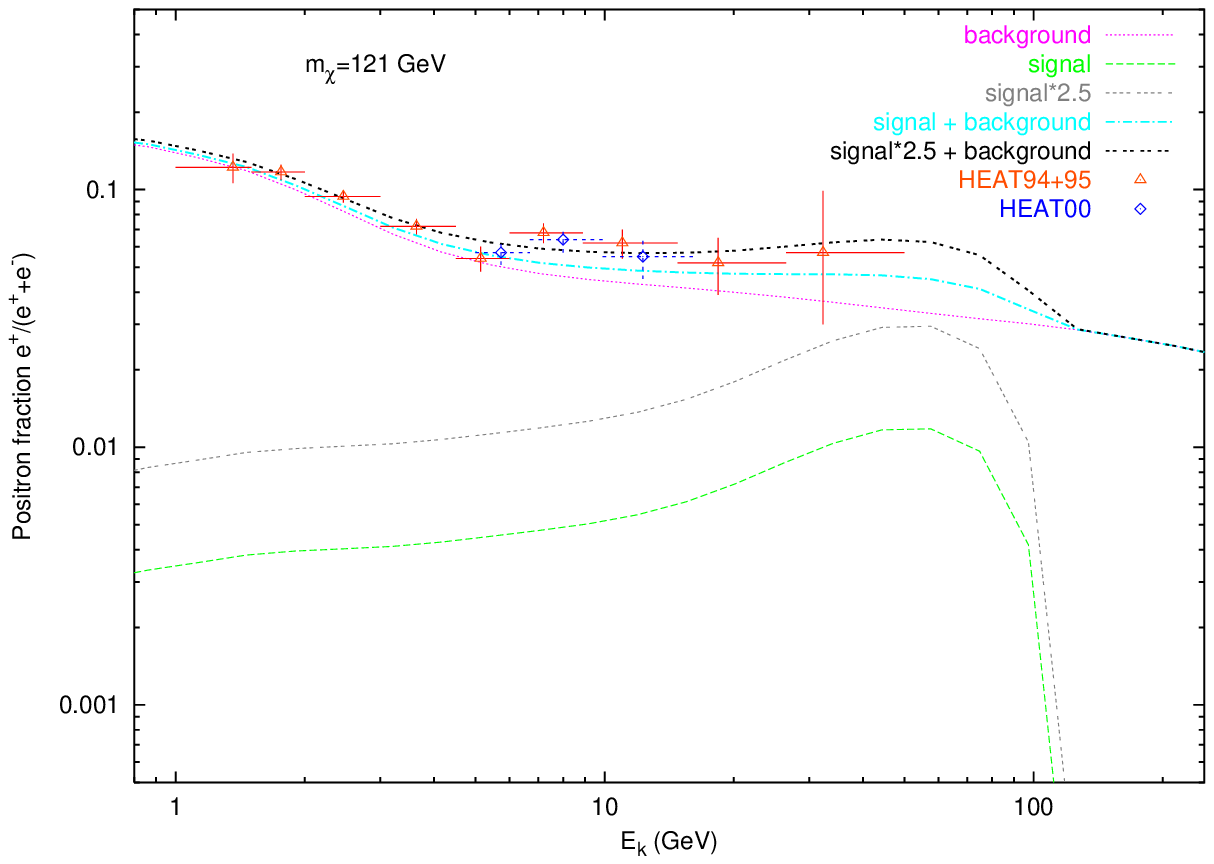}}
\caption{Positron fraction $e^+/(e^++e^-)$ calculated in this work. The
left panel shows the result of a Moore DM profile 
without substructures. The right one is the same as the
left but including the 
subhalos with $\gamma=1.7$ and $\rho_{\rm max}=2\times 
10^{19} M_\odot/kpc^3$. Data points are HEAT measurements in the 
three fights in 1994, 1995 and 2000 \cite{Barwick97,Coutu01}.}
\label{ratio}
\end{figure}

\begin{figure}[htb]
\resizebox{13.cm}{!}{\includegraphics{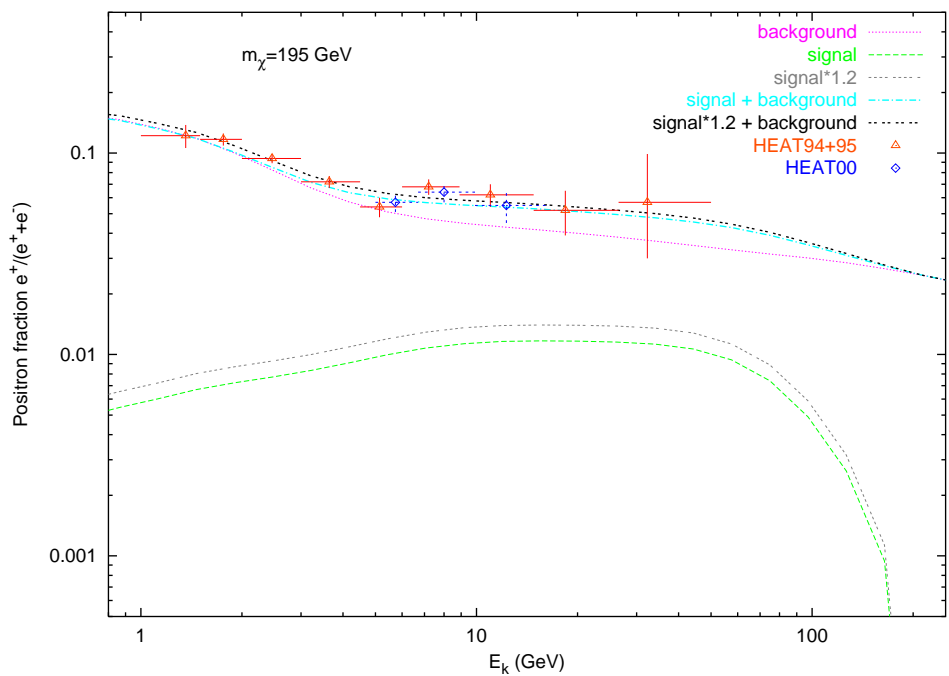}}
\caption{The same as Fig. \ref{ratio} but for neutralino with mass
$m_{\chi}=195GeV$.}
\label{ratio195}
\end{figure}

To give a consistent Galactic positron flux we calculate the background 
(secondary) positron fraction generated by interaction between primary
cosmic-rays and interstellar gas in the same propagation model using
GALPROP. In the Figures \ref{ratio} and \ref{ratio195} we give the 
background positron fraction. Our results give better description to 
HEAT data compared with the adapted background fraction in \cite{Baltz99,
Kane02,Baltz02,Cumberbatch06}. Therefore the new results requries a
smaller ``boost factor'' to the DM annihilation signals.

Adding the positron flux coming from DM annihilation to the secondary
positrons we get the total positron flux. The
positron fraction $e^+/(e^++e^-)$ is plotted in Fig. \ref{ratio}
for $m_\chi =121$ GeV. We can see from this figure that the expectation
reproduces the measurements relatively well if DM subhalos are taken
into account. The model with subhalos can contribute several times
more than which without subhalos. The best fit to the HEAT data requries
``adjustment factor'' of $9.2$ and $2.5$ for signals from DM annihilation
without and with subhalos respectively.

We don't think the discrepancy of a factor $2.5$ is serious since there
are large uncertainties in calculating the positron fluxes. 
Firstly there are uncertainties for the positron propagation, as
shown in Figure \ref{propmodel}. Secondly the random distribution
of DM subhalos may have large variance as discussed in \cite{Lavalle06}
since positrons lose energy quickly and have large fluctuations. The
variance may lead to large positron flux due to accidence, while in this
work we just give the average positron flux. Thirdly we can choose
SUSY model which produces more positrons, as shown in Figure \ref{ratio195}
for a model with $m_{\chi}=195GeV$. The best fit to HEAT data requries
a factor of $1.2$ for this model. 
It is remarkable that without introducing a large ``boost factor'' nor
the nonthermal production of neutralinos as done in \cite{Kane02}
we could explain the HEAT data quite well.

It is interesting to notice that at about 8$GeV$, HEAT measurements show
some fine structures of the positron fraction, which cannot appear in our
calculation. Maybe there are some specific sources or some specific physical
process to generate positrons at this energy. However, the errors of
the data are too big to give a definite assertion of this property. If it is
confirmed by further experiments such as AMS-2, there should be
deeper studies about the models we used here.

In summary, in this work we give a consitent and detailed calculation of
the Galactic positron flux. We recalculate the background positron flux
from cosmic-ray collisions with the interstellar gas in a realistic 
propagation model GALPROP. The results soft the discrepancy between
data and expectations. For the primary positrons from DM annihilation
we take the enhancement of subhalos into account. The result shows that
the HEAT data can be explained by the two components without introducing
``boost factor'' or nonthermal production of DM. To account for the excess
of positron flux extending subhalo masses to $10^{-6}M_{\odot}$ and 
requiring a cuspy profile as $\gamma=1.7$ are necessary.

\ack
We thank Igor V. Moskalenko for great help on using the package GALPROP.
This work is supported in part by the NSF of China under the grant
No. 10575111, 10120130794 and supported in part by the
Chines Academy of Sciences under the grant No. KJCX3-SYW-N2. 


\end{document}